# The derivative expansion of the renormalization group

Tim R. Morris[a]*

[a]Physics Department, University of Southampton,
Southampton, SO17 1BJ

By writing the flow equations for the continuum Legendre effective action (a.k.a. Helmholtz free energy) with respect to a particular form of smooth cutoff, and performing a derivative expansion up to some maximum order, a set of differential equations are obtained which at FPs (Fixed Points) reduce to non-linear eigenvalue equations for the anomalous scaling dimension $\eta$. Illustrating this by expanding (single component) scalar field theory, in two, three and four dimensions, up to second order in derivatives, we show that the method is a powerful and robust means of discovering and quantifying non-perturbative continuum limits (continuous phase transitions).

In this short review I want to stress what I believe to be the essential steps involved in the scheme briefly described above[2,3], and compare and contrast other analytic approximation approaches to the RG (Renormalization Group). Particularly in the two dimensional example[4], the method is seen to be in several ways more powerful in fact than any other approximate approach, analytic or otherwise. This is also an excellent opportunity to refer to some papers which I was unaware of at the time of writing refs[1–3].

The first point to make is that any continuum approximation method, if it is to be of general applicability, must be formulated within the framework of Wilson's RG[5]. This is true, even if one looks only at perturbation theory to all orders[1], but is particularly obvious when the theory at the UV cutoff scale $\Lambda_0$, where the bare action $S_{\Lambda_0}$ is formulated, is also strongly interacting. In this case the usual power counting arguments, to determine which are the renormalizable (i.e. relevant) interactions, breaks down and the definition of the Lagrangian itself becomes part of the calculational problem. Breaking this circularity was one of the many successes of RG viewpoint. In terms of this one should think of the space of all possible local interactions. The theory is then renormalizable if it is describable in terms of self-similar evolution in some subspace of parameters under the flow of some low energy scale $\Lambda \ll \Lambda_0$.

*This work was supported by a CERN fellowship and PPARC Advanced fellowship.

In the way this is usually formulated one thinks of an effective action $\Gamma_\Lambda$ designed to work with a UV cutoff $\Lambda$ in such a way that it reproduces exactly the same physics as one obtains with $S_{\Lambda_0}$ (and one often thinks of rescaling the cutoff back to its original size after the "blocking") but these concepts are not really necessary.

The next question to ask is: What are the possible approximations in this approach? The obvious approximation is to truncate the space of interactions to just a few operators. This direction has been extensively pursued over the years, especially within real space RG[6], and the results can sometimes be very accurate – and sometimes not. Although by no means as sophisticated yet, approximations by truncations of the continuum RG also have a long history[5,7]. However, if the field is strongly interacting, the couplings of all the interactions $\phi^{2n}$ will be more or less the same size, so these methods need a lot of care. (Reasons for some limited accuracy are given in ref.[3]). To have a chance of finding a robust method of general applicability we must keep at least infinitely many such operators.

We will do so by using a $\partial$E (Derivative Expansion) up to some maximum order. (Maybe there are other ways). Thus to $O(\partial^2)$ for example we write

$$\Gamma_\Lambda = \int d^D x\, V(\phi) + \frac{1}{2}(\partial_\mu \phi)^2 K(\phi) \qquad (1)$$

with general functions $V$ and $K$, and ignore $O(\partial^4)$

terms and higher. This expansion seems particularly natural, and one might hope that it has general applicability in the sense that if the higher derivative terms are not small, then the description in terms of a single scalar field is probably itself inappropriate and indicates that other degrees of freedom should be introduced.

So far we have dealt with rather unsubtle requirements, but there is another requirement, which determines the form of the cutoff, and to highlight this it will help to consider the simplest sort of self-similar evolution, namely when $\Gamma_\Lambda$ is actually *invariant* under lowering $\Lambda$, i.e. FPs (Fixed Points) of the RG. Since these occur at zeroes of $\beta$ functions, we generally expect only a few such FPs, and that at each FP all parameters will be dynamically determined. Independence of $\Lambda$ implies scale invariance, and thus a massless theory. As is well known, this scale invariance is realised only if one assigns the field the dimension $[\phi] = d_\phi = \frac{1}{2}(D - 2 + \eta)$, where $D$ is the space-time dimension and $\eta$ is the anomalous dimension. $\Lambda$ must be introduced in a way which is equivalent to an IR cutoff $C(p,\Lambda)$[4]. Introducing it into the partition function as

$$Z[J] = \int \mathcal{D}\phi \; \exp\{-\frac{1}{2}\phi.C^{-1}.\phi - S_{\Lambda_0}[\phi] + J.\phi\} \quad (2)$$

it is then easy to write down a flow equation: $\frac{\partial}{\partial \Lambda} Z[J] = -\frac{1}{2} \frac{\delta}{\delta J} . \frac{\partial C^{-1}}{\partial \Lambda} . \frac{\delta}{\delta J} Z$. Now we rewrite all quantities in terms of dimensionless quantities, using $\Lambda$, so that everything will be independent of $\Lambda$ at a FP. (From (2) by dimensions, we require $[C] = \eta - 2$). It is also very helpful to rewrite it as an equation for the Legendre effective action $\Gamma[\phi]$. There are a number of reasons for this[1, 2,4]; I will mention one later. Having done this we obtain an equation for $\Gamma[\phi]$ with the following features[2]: It has $\delta^2\Gamma/\delta\phi^2$ (as a consequence of the two $J$ derivatives), and $\eta$ appears explicitly now as a parameter in the equations.

Consider what happens when we look for FPs at lowest order in the $\partial$E, $O(\partial^0)$, where we just set $K = 1$ in (1). Since this corresponds to throwing away all momentum dependent corrections, it is clear that $\eta = 0$ in this approximation. Substituting this into our flow equation, it follows from the previous description that one obtains a (non-linear) second order (ordinary) differential equation for $V(\phi)$. Such an equation has generically a two parameter set of solutions. If these all looked acceptable it would indicate that we would need further information and cannot just examine the FPs directly in this way. Fortunately some magic occurs. It turns out that (generally) very nearly all solutions are unacceptable because they are singular at some finite real $\phi = \phi_c$ [2–4]. The few non-singular solutions are approximations to the physical FPs. Effectively $O(\partial^0)$ equations have been considered before[8,9] with similar conclusions[8,10–12]. If we fix $V'(0) = 0$, considering only $\phi \leftrightarrow -\phi$ invariant theories, then the value of $\phi_c$ can be plotted against, say, $V(0)$. In this way we can scan through the *infinite dimensional space of all possible potentials* $V(\phi)$, looking for continuum limits ($\phi_c = \infty$) [3]. Clearly this is much more than is possible with other approximate methods. However, only the expected FPs have been found at this level. In particular only the Gaussian FP in 4 dimensions[11,3], this and the Wilson FPs in 3 dimensions[10–12,2,3], and in 2 dimensions only critical sine-Gordon models[4]. There are very strong conjectures for the existence also of an infinite set of multicritical FPs[13] in 2 dimensions, but this requires $\eta > 0$. I will come back to these, but first I want to show that there is another way of seeing that there are only a few FP solutions: On the one hand $V'(0) = 0$, and on the other hand, if the cutoff $\Lambda$ drops out we have, in the limit $\phi \to \infty$, $V(\phi) \propto \phi^{D/d_\phi}$, by dimensions. This gives us now the needed two boundary conditions which indeed are satisfied by the FP solutions (providing $d_\phi \neq 0$).

It is useful to extend this argument to $O(\partial^2)$ and higher. Thus consider the FP equations one obtains now by using (1) with $K$ a fully fledged function. Just as before, the structure of the flow equations means that we get second order ordinary differential equations, now a coupled pair of them. Now we expect two extra parameters, but we also have now two more constraints, coming from symmetry and dimensional arguments on $K$. In this way, to all orders of the $\partial$E, the coefficient functions are determined for each FP. However we have been ignoring $\eta$, which can now be non-zero, and if the method is to work at all it must also get determined somehow. It gets determined in the exact ($\equiv O(\partial^\infty)$) equations via a reparametrization invariance[14] of the flow equations, involving a rescaling of the field[14–16]. The point is that this extra invariance means that we can fix an extra condition, which now overconstrains the equations, turning them into (non-linear) eigenvalue equations for $\eta$. (E.g. use it to normalise the kinetic term as $K(0) = 1$, as in fact was done implicitly at $O(\partial^0)$). The problem is that this invariance is generally broken by the approximation scheme[15], including the $\partial$E[17], and one is reduced to heuristics for trying to find a 'best' value for $\eta$[15–17]. Actually there is a, probably unique, form of cutoff for which the $\partial$E *preserves* this invariance: $C(q^2) \propto q^{2\kappa}$, for some integer $\kappa > D/2 - 1$ [2]. This only cuts off the Legendre equations, and only if this inequality is satisfied. It works because the invariance is then just another scaling symmetry with a (non-physical) set of scaling dimensions. Clearly the $\partial$E will converge best for the smallest value of $\kappa$[1,2].

For the operator spectrum, nothing new is needed. Linearising the flow equation around the FP solutions, we have again precisely the right number of boundary constraints, from symmetry, dimensional analysis and the normalisation allowed by linearity[2].

It remains to state that the results of this method are so far very encouraging. In particular to $O(\partial^2)$ one still finds only Gaussian and Wilson FPs in three dimensions, but now with estimates for critical exponents which are within six times the error of the best present estimates[2]. In two dimensions, a constrained search finds only the conjectured $\eta > 0$ multicritical points[4]. For the first ten such scalar field theories (critical, tricritical, quadricritical, ..., undecimcritical) the exponents and dimensions of up to the first ten operators agree within 33% with expectations[13] over two orders of magnitude, with a weak *improvement* with increasing multicriticality and dimension. (We may also have found "shadow operators"[4]). These facts are particularly significant, since *all* other present approximate nonperturbative methods, lattice Monte Carlo, resummations of weak or strong coupling perturbation theory and the $\epsilon$ expansion, become rapidly impracticable with increasing multicriticality or dimension[4].

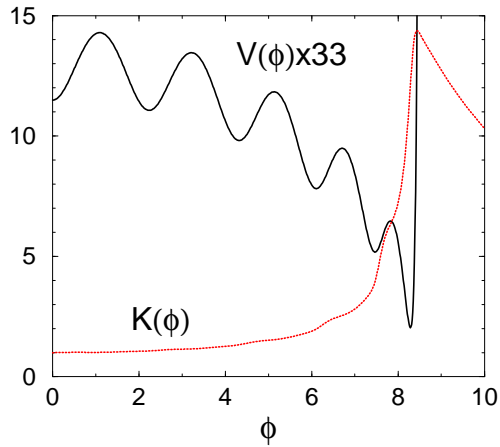

The two dimensional undecimcritical point. $V(\phi)$ multiplied by 33 for display purposes.